\documentclass[letter]{aa} 

\usepackage{graphicx}
\usepackage{txfonts}
\usepackage{array}
\begin{document}

   \title{The first radial velocity measurements of a microlensing event: \\no evidence for the predicted binary\thanks{Based on observations made with ESO Telescope at the Paranal Observatory under program ID 092.C-0763(A) and 093.C-0532(A).}}


   \author{I. Boisse\inst{1}
           \and
A. Santerne\inst{2}
           \and
J.-P. Beaulieu\inst{3}
           \and
W. Fakhardji\inst{1}
           \and
N.C. Santos\inst{2,4}
           \and
P. Figueira\inst{2}
       \and
S. G. Sousa\inst{2}
\and
C. Ranc\inst{3}
          }

   \institute{Aix Marseille Universit\'e, CNRS, LAM (Laboratoire d'Astrophysique de Marseille) UMR 7326, 13388, Marseille, France \\
              \email{isabelle.boisse@lam.fr}
         \and
             Instituto de Astrof\'isica e Ci\^encias do Espa\c{c}o, Universidade do Porto, CAUP, Rua das Estrelas, 4150-762 Porto, Portugal  
             \and
               CNRS, Universit\'e Pierre et Marie Curie, UMR 7095, Institut d'Astrophysique de Paris, 98 bis boulevard Arago, F-75014 Paris, France
               \and
             Departamento de F\'isica e Astronomia, Faculdade de Ci\^encias, Universidade do Porto, Rua do Campo Alegre, 4169-007 Porto, Portugal 
             }

   \date{Received ; accepted }

 
  \abstract
   {The gravitational microlensing technique allows the discovery of exoplanets around stars distributed in the disk of the galaxy towards the  bulge. However, the alignment of two stars that led to the discovery is unique over the timescale of a human life and cannot be re-observed. Moreover, the target host is often very faint and located in a crowded region.  These difficulties hamper and often make impossible the follow-up of the target and study of its possible companions.     
   
    Gould et al. (2013) predicted the radial-velocity curve of a binary system, OGLE-2011-BLG-0417, discovered and characterised from a microlensing event by Shin et al. (2012). We used the UVES spectrograph mounted at the VLT, ESO to derive precise radial-velocity measurements of OGLE-2011-BLG-0417. To gather high-precision on faint targets of microlensing events, we proposed to use the source star as a reference to measure the lens radial velocities.
   We obtained ten radial velocities on the putative V=18 lens with a dispersion of $\sim$100~m\,s$^{-1}$, spread over one year.  Our measurements do not confirm the microlensing prediction for this binary system. The most likely scenario is that the assumed V=18 mag lens is actually a blend and not the primary lens that is 2 mag fainter.
Further observations and analyses are needed to understand the microlensing observation and infer on the nature and characteristics of the lens itself.
   }

   \keywords{planetary systems --
                radial-velocity --  microlensing -- OGLE-2011-BLG-0417}
\maketitle
%

\section{Introduction}

   Different exoplanet detection methods (radial velocity, hereafter RV, stellar transits, direct imaging, pulsar timing,
astrometry, and microlensing) are currently used to probe different populations of planets over a wide
range of orbital radii, masses and host types. 
To date, around 25 exoplanets have been discovered via microlensing, and roughly as many await publication (Beaulieu, priv com).
These numbers are relatively modest compared
with that discovered by the RV method or by the Kepler satellite. However, microlensing probes a domain of the
parameter space (host separation vs. planet mass) which is often not accessible currently to other methods. The detection of cold planets down to a few Earth masses (Beaulieu et al. 2006) or the observation of free floating planets put the
planetary formation scenario to the test (Cassan et al. 2012). On the other hand, microlensing events have the weaknesses of not being repeatable, and to focus on faint stars on crowded fields that are difficult to characterise precisely. A reobservation of the system to get further parameter characterization is often very difficult.

From the observed microlensing light curve, the first determined parameters are the mass ratio and the sky-projected 
angular separation of the system. Additionnal effects, parallax, xallarap, terrestrial parallax, finite source effects, detection of the light coming from the lens (thanks to high angular resolution), or Bayesian analysis are used in order to derive the physical
parameters of the different planetary systems and to proceed further in the analysis. 
In this context, Skowron et al. (2011) showed that the deformation of the microlensing light curve can be used to constrain all the orbital parameters of a binary lens system.

 Shin et al. (2012) presented the microlensing event OGLE-2011-BLG-0417 and modelled it as due to a binary lens system. 
  The source star is a K3 red giant located in the galactic bulge at 8 kpc with I$_{\mathrm source}$=16.74 (V$_{\mathrm source}$=19.42). 
They identified the blended light as the primary lens, (I$_{\mathrm lens}$=16.30, V$_{\mathrm lens}$=18.23), which would make it 
one of a few case where the lens primary is significantly brighter than the source star. 
Gould et al. (2013) adopted the new calibration of Nataf et al. (2013) for the Bulge giant and revised the initial error budget. Located at $0.95\pm0.06$~kpc, the lens binary is composed of a primary star of $0.524 \pm 0.036$ M$_\odot$, orbited by a M dwarf of $0.153 \pm 0.011$~M$_\odot$. 
They showed that this event could be tested by RV measurements and published revised Keplerian parameters (reported here in Table~1). With a large expected RV amplitude, of 6.4~km\,s$^{-1}$, and with an eccentricity of 0.069, this system can be detected using only relatively low-precision instruments, benchmarking the microlensing detection.

In this letter, we present the first radial velocity observations of this microlensing target, using the UVES spectrograph.

   \begin{table}
      \caption[]{RV parameters derived by Gould et al. (2013) from the analysis of the microlens event.}           
\label{table_paramK}      
\centering                          
\begin{tabular}{c c c c c c}       
\hline\hline                 
 & K & P & e & $\omega$ & T$_{peri}$ \\ 
 & km\,s$^{-1}$ & yr &  & deg & HJD\\
\hline                        
 Value &  6.352  & 1.423 & 0.688 & 341.824  & 5686.344 \\   
Error &   0.340 & 0.113 & 0.027 & 2.655   &  6.960 \\
\hline                                   
\end{tabular}
\end{table}
%
%


\section{Observations}

We obtained a total of 9 hours of observations with the UVES cross-dispersed echelle spectrograph (Dekker et al. 2000) mounted on the VLT in P92 and P93 to measure the RV variations of the OGLE-2011-BLG-0417 binary.  Ten spectra were acquired between October 2013 and September 2014.

We used the two arms of the spectrograph in parallel with a dichroic beam splinter, the standard mode DIC-1 (390+580), with a wavelength coverage of 326-445 and 476-684~nm. It probes a domain where the late-K dwarf lens emits sufficient flux and where the spectra is not strongly polluted by telluric lines. 
The exposure time was set to one hour in order to reach a signal-to-noise ratio (SNR) of $\sim 20$ at 550~nm. We used a slit of 1 arcsec that gives a spectral resolution of 40000, sufficient to resolve the lines and calculate RV without loosing  too much light due to slit losses. Due to the faintness of the target, the guiding was done with the red camer, the slow readout mode of the CCD was used and we requested a seeing no larger than 1''. Still, a slightly larger seeing than the width of the slit leads to some flux loss, but this is compensate in RV precision by a gain in the stability of the illumination of the slits of the spectrographs (Boisse et al. 2010). The log of the observations is given in the online Table~3. All the measurements were kept for the analysis.

UVES is not stabilised in pressure and temperature. An important RV drift of the zero point is expected to be present and depend on the temporal evolution of the two parameters. 
To minimize this effect, a thorium-argon calibration was done before and after the scientific observation.

The target star being in the Galactic bulge, the field is densely crowded. We then fixed the angle of the slit with the sky as shown in the top of Fig.~1 in order to minimise the pollution of contaminant star inside the slit.

%
   \begin{figure}
   \centering
   \includegraphics[width=8cm]{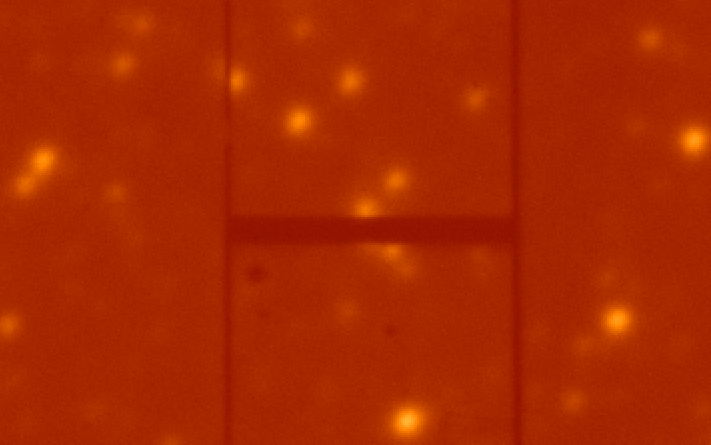}
   \includegraphics[width=8cm]{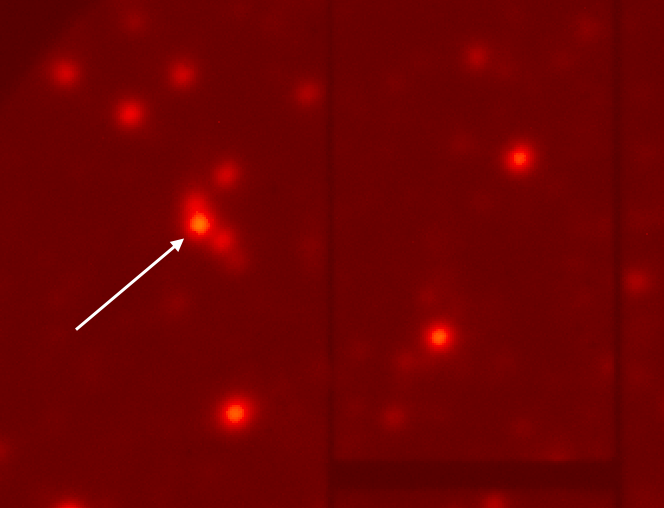}
      \caption{Images of the slit-viewer camera of UVES red-arm showing the crowding of the field. The black horizontal shadow is the slit opened at 1''. On the top, position that we chose when the slit is on the target, OGLE-2011-BLG-0417. On the bottom, the target is decentered (top left of the slit). North is up and East is on the right.       }
         \label{Fig_field}
   \end{figure}
%

\section{Data reduction }

We used the Reflex ESO pipeline to reduced the spectra (Freudling et al. 2013). We first corrected them from cosmic rays with a $5\sigma$-clipping process.  
We then computed the weighted cross-correlation functions (CCF) of the spectra with a template K5 mask (Pepe et al. 2002). The CCF of the co-added spectrum is plotted in Fig.~2. The source and the lens are superimposed. 
Their RV values are different by several tens of km.s$^{-1}$, as expected since the lens is in the Galactic disk whereas the source is in the bulge. 
The observation of two significant components shows that the observed target was indeed OGLE-2011-BLG-0417.

 We compared the correlations obtained for the three detectors (one in the blue and two in the red arm) to disentangle the source star from the lens. The variation as a function of wavelength of the relative contrast of the CCF is equivalent to flux ratio. So, the bluer source would show a deeper contrast in the blue wavelength domain. With a $(V-I)$$_{\mathrm lens}$=1.93 and a $(V-I)$$_{\mathrm source}$=2.68, the source is indeed expected to be bluer than the lens. On the contrary, the redder lens shows a higher contrast in the red. The source also presents a deepest CCF as presumed for a lower log~$g$ (giant star).
 
We checked with the slit-viewer camera that no other source entered the 1 arcsec slit (see bottom of Fig.~2). On three observations, however, the seeing increased or the slit drifted from the pointed target (see the online Table~3). As a consequence, two of the close-by stars entered the slit. In the individual spectra, two components with low contrast appeared in the CCF profiles. However, they are well separated from the lines of the source and the lens. 
We tested that the CCF width of the source and the lens are constant within the error bars for our 10 measurements, and hence that the measured RV are not polluted by contaminant stars. We note, however, that the presence of contaminants would in principle increase the RV dispersion. 

To be put in the same RV heliocentric reference, each CCF profiles were corrected in wavelength from the barycentric earth radial velocity (BERV) values given by the UVES pipeline.  The CCF profile were then fitted with a two-Gaussian model to derive the RV of the lens and the source. A third and fourth Gaussians were used when contaminant stars were visible. For each observation, we calculated two RV values, one for each of the wavelength calibration secured before or after the science exposure: RV$_{\mathrm before}$ and RV$_{\mathrm after}$ (respectively). The derived RV is the average between the two, which assumes that the instrumental drift is nearly linear between the two calibrations. 
 We measured the spectrograph RV drift to be from 15  up to 400~m\,s$^{-1}$, within one hour. 
Using the median exposure as calculated from the photometer count of the detectors and weighting by the spectral information between the blue and red arms does not change the result above a few~m.s$^{-1}$. 
 This correction of the drift allows us to reach  root mean square (RMS) for each of the measured stars of $\sim$800~m.s$^{-1}$. 
 
A zero-point drift of the spectrograph may have not been corrected by the thorium-argon calibrations. The RV is moreover dispersed by another source of noise due to changes in the illumination of the slit for the different spectrum (and so different point spread functions). The long exposure time (one hour) was expected to average the small movement of the target in the slit. But, the seeing was often significantly smaller than the slit (see the online Table~3). At this stage, the main contribution to the dispersion should come from this effect, however, difficult to measure precisely. 

We then computed a telluric mask from O$_{2}$ lines (Figueira et al. 2010) and cross-correlate it with each spectra. The obtained CCF was fitted by a Gaussian to derive a zero-point RV value. Only 20 O$_{2}$ lines could be fitted in the spectral domain, and the derived RV present an error of $\sim$200~m.s$^{-1}$. When we further corrected the stellar RV values from the zero-point drift as calculated from the telluric lines, the RMS decreases to $\sim$250~m.s$^{-1}$. 
We plotted in Fig.~3, the RV of the source and the lens. The two stars are not gravitationally bound and their RV variations are dominated by the same instrumental systematics. The main contribution comes from the error on the telluric correction. 

We then decided to use the RV from the source as a reference to measure the lens RV. It leads to the $\Delta$RV value. This could be done because the flux of the two stars are blended in the spectra and followed the same path in the spectrograph. 
By doing that, the systematics observed in both stars are canceled out.

The photon-noise uncertainty was estimated from the empirical calibration of Bouchy et al. (2005) on UVES/FLAMES (see also Loeillet et al. 2008). That leads to a mean error bar of $\sigma_{\mathrm pn}\sim$50~m.s$^{-1}$ on the individual targets, and we took their quadratic sum for the $\Delta$RV photon noise error. Another source of noise comes from the drift of the instrument during the exposure. Considering this drift is linear between the calibrations and assuming that the value of the RV has a Gaussian distribution that is within the two calibrations with a probability of 99.9\%, we calculated an error of $\sigma_{\mathrm calib}\,=\,(RV_{\mathrm before}\,-\,RV_{\mathrm after})/ (2 \times 3)$. 
The RV of the source, the lens and the $\Delta$RV, as well as their error bars are reported in the online Table~2. The reported errors are the quadratic sum of  $\sigma_{\mathrm pn}$ and $\sigma_{\mathrm calib}$. 
The mean error is $\sim$110~m.s$^{-1}$ for the individuals RV and 65~m.s$^{-1}$ on the $\Delta$RV.


%
%
   \begin{figure}
   \centering
   \includegraphics[width=\hsize]{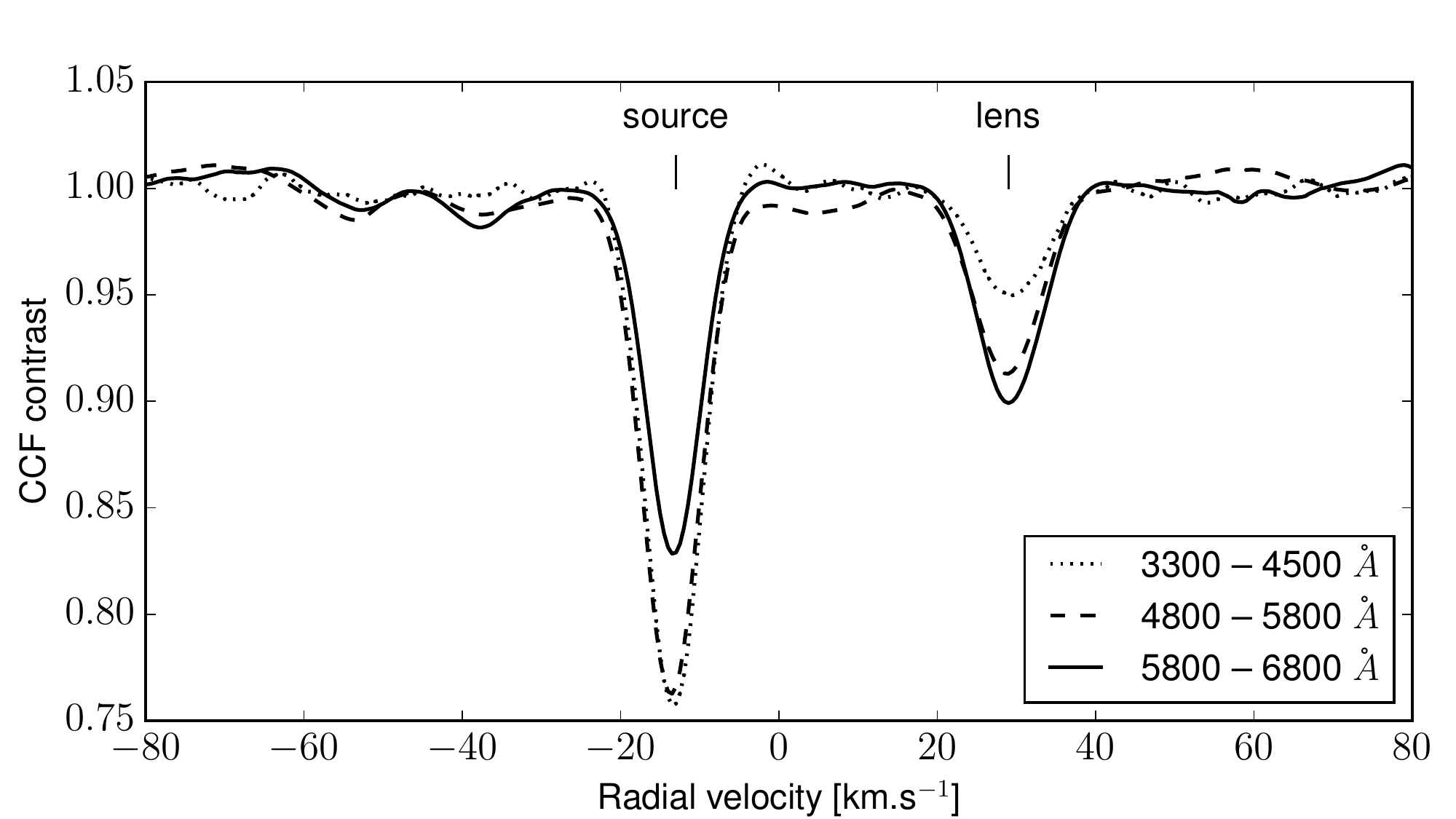}
      \caption{CCF compute on three bandpasses of the sum of all the spectra, allowing to identify the thinner and bluer line 
      to be the giant source and the other, the late-K dwarf primary of the binary lens.}
         \label{Fig_CCF}
   \end{figure}
%

\section{Data analysis and results}

The ten secured RV present an RMS of 94~m\,s$^{-1}$, with no significant variations. We plotted them in Fig.~4 with the best model from Gould et al. (2013). There is a clear disagreement between the observations and the predicted model. Using the PASTIS validation software (Diaz et al. 2014), we estimated the probability of this predicted model. We modelled a Keplerian orbit with normal priors that matched the values and uncertainties reported by Gould et al. (2013). We used a uniform prior for the systemic RV as well as for an extra RV jitter. We ran 20 Markov Chain Monte Carlo analysis of $3.10^{5}$ iterations each. We repeated the same analyses with an opposite sign for the RV amplitude (the sign of the RV curve is not determined by the microlensing prediction) and for a no-variation scenario. After thinning and merging the chains, we ended with about 10 000 independent samples of the posterior distribution.

The best-fit model is superimposed in grey in Fig.~4. The residuals exhibit a RMS of 340~m\,s$^{-1}$, hence three times larger than the dispersion of the data. Moreover, this best-fit model departs from the prediction joint confidence interval by 3.7 sigma (a priori probability of $\sim2.10^{-4}$). We then estimated the statistical evidence of the Gould et al. (2013) prediction, using the method described in Tuomi \& Jones (2012). We found a probability of $7.10^{-8}$ or $2.10^{-7}$ depending if we consider a RV amplitude that is positive or negative (respectively). Therefore, our spectroscopic observations unambiguously reject the microlensing prediction of Gould et al. (2013) for this binary system. 

We decided to double check if the bright blend (I$_{\mathrm blend}$ = 16.29, V$_{\mathrm blend}$ =18.23) is the primary component of the lensing system as claimed by Gould et al. (2013). We adopt a set of isochrones (Girardi et al. 2012) with ages in the range 1-10 Gyr for a solar metallicity and the distance modulus of 950 pc.  We find a G8 star with a mass of  $\sim$0.82~M$_\odot$, almost 2 magnitudes brighter than the primary lens. The bright blend cannot be the primary lens of the microlensing event OGLE 2011-BLG-417.   We will revisit the system with high angular resolution to see if the bright blend could be a distant companion to the lensing system or a chance alignement on the line of sight of the source.

   \begin{figure}
   \centering
      \includegraphics[width=8.9cm]{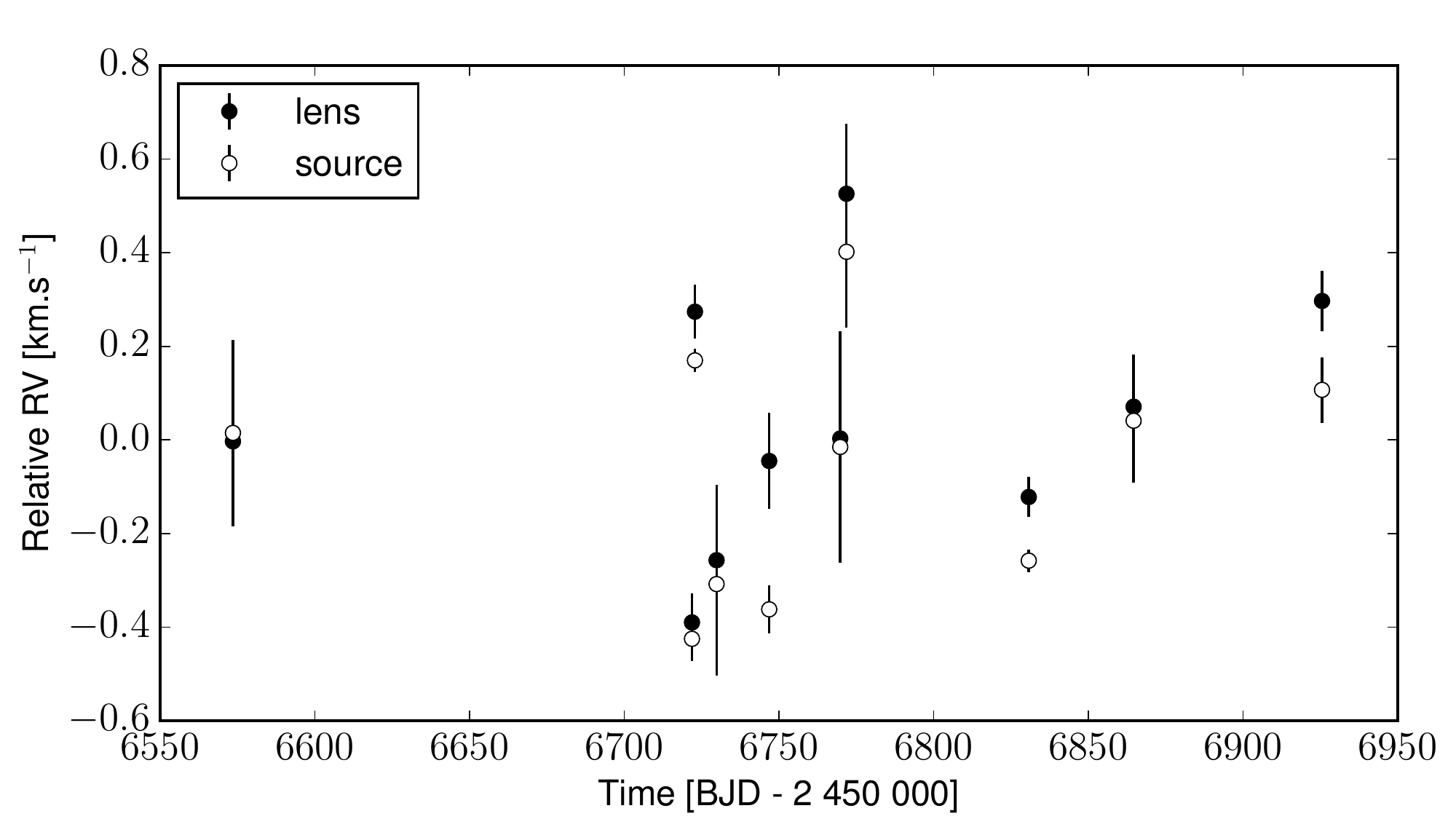}
      \caption{UVES RV of the source and the lens. The velocities of the two stars share the same instrumental systematics.}
         \label{Fig_RV_stars}
   \end{figure}
%

%
%
   \begin{figure*}
 \centering
  \includegraphics[width=15cm]{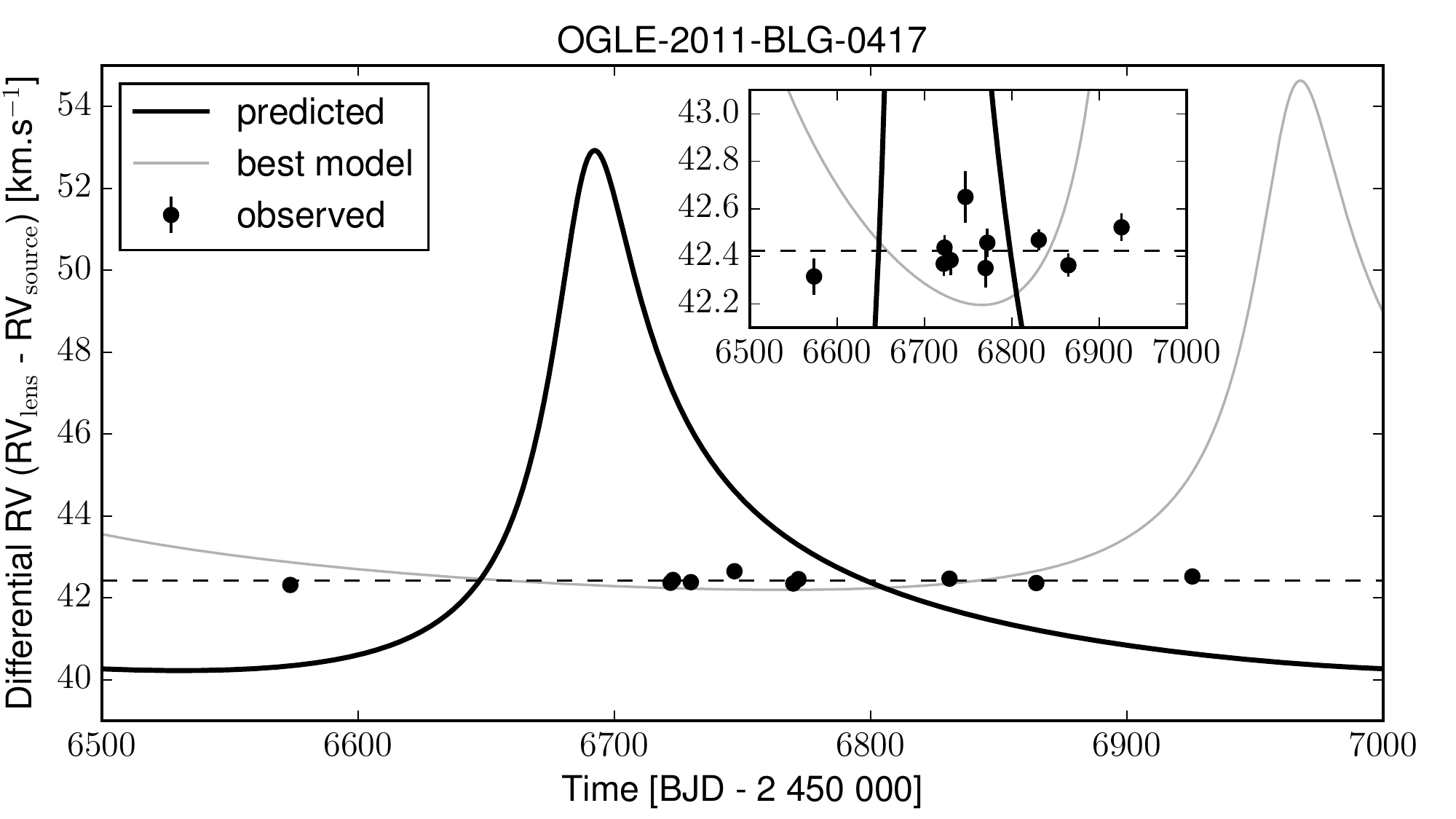}
     \caption{UVES RV of OGLE-2011-BLG-0417 (dots) with the predicted model
from Shin et al. (2012) and Gould et al. (2013) (black line). The systemic was fixed to the median value of the observations. The best-fit model that departs from the prediction by 3.7 sigma, is superimposed in grey. The observations reject the modelled Keplerian 
at a level greater than $2.10^{-7}$ (see text). The insert is an enlargement around the observations.}
         \label{Fig_RV_result}
   \end{figure*}


\section{Conclusions}
We used the UVES spectrograph to derive Doppler measurements of the reflex motion of the primary component of the predicted binary lens OGLE-2011-BLG-0417. The lens, composed of a  late-K dwarf orbited by a M dwarf, is brighter (V=18) than the microlensing source (V=19.3). The huge semi-amplitude of $\sim6.35$~kms$^{-1}$ of the predicted eccentric orbit should have been easily detected with the RV precision reported her.

 Our ten measurements with a dispersion of 94~ms$^{-1}$ and a mean error bar of 65~ms$^{-1}$ do not confirm the microlensing analysis.  These are the first published RV measurements on a microlensing target. 
We are led to believe that an error on the analysis of the microlensing event is the reason of the discrepancy. A quick look at the photometric data of the microlensing event seems to indicate that the brightness of the lens has been overestimated. The most likely scenario is that the bright blend of the microlensing light curve is not the light from the primary lens
in contradiction with the prediction of Gould et al. (2013). As a consequence,
the RV modulation of the lensing system could not be detected, because the primary is 2 magnitude fainter. It is not clear yet if the bright blend is a distant companion to the OGLE 2011-BLG-417 system or a chance alignement.
 A more complete reanalysis of the system taking into account spectroscopic and photometric data will be performed in due time, but this is outside the scope of this letter. 

UVES allowed us to reach a RV precision of 100~m\,s$^{-1}$ on a target of $18^{\mathrm th}$ magnitude in V in 1 hour-time
integration. 
This precision was made possible thanks to the fact that the source star is observed simultaneously
with the binary microlens and can serve as calibration. The technique is similar to differential photometry, but in the spectral domain. 

  With a modest allocation of telescope time (9 h), it would have been possible to characterised with RV, a binary system
detected by microlensing.  
This has strong implication for the modelling of the microlensing observations. This shows that spectroscopic follow-up observation of microlensing events is possible with large telescopes. 
Understanding the reasons of the discrepancy on this event will help to improve the characterisation of microlensing systems, already detected or that will be detected with the K2, WFIRST and Euclid space missions. 
In the coming years, using future ESO facilities such as ESPRESSO @ VLT, or HIRES @ E-ELT it will be possible 
to perform such measurements on planetary systems detected by microlensing such as OGLE-2007-BLG-109 (Gaudi et al. 2008) and OGLE-2012-BLG-0026 (Han et al. 2013), once it has been confirmed via high angular resolution observation that there is no strong contamination by a blend, for example thanks to high angular resolution observations.

\begin{acknowledgements}
     
     We thanks the UVES operation astronomers that performed these observations and Christian Hummel for his help in the preparation of the p2pp process.  I.B. thanks A.S. for his enthusiastic work. Between the first and last measurements, both of us became parent. What a great change ! A.S. warmly thanks Rodrigo F. D\'iaz and Jos\'e-Manuel Almenara for their substantial contribution in the development of the PASTIS software. 
      A.S. is supported by the European Union under a Marie Curie Intra-European Fellowship for Career Development with reference FP7-PEOPLE-2013-IEF, number 627202. This work was supported by Funda\c c\~ao para a Ci\^encia e a Tecnologia (FCT) through the research grant UID/FIS/04434/2013. P.F., N.C.S., and S.G.S. also acknowledge the support from FCT through Investigador FCT contracts of reference IF/01037/2013, IF/00169/2012, and IF/00028/2014, respectively, and POPH/FSE (EC) by FEDER funding through the program Programa Operacional de Factores de Competitividade -- COMPETE. P.F. further acknowledges support from FCT in the form of an exploratory project of reference IF/01037/2013CP1191/CT0001.
      
      \end{acknowledgements}


\Online

\begin{table}
      \caption[]{RV measurements and their associated 1-$\sigma$ error bars.}           
\label{table_RV}      
\centering                          
\begin{tabular}{|p{1.5cm}| p{1.1cm} p{0.7cm} p{0.8cm} p{0.74cm} | p{0.8cm} p{0.74cm}|}       
\hline\hline                 
BJD$^{(1)}$ & RV$_{source}$ & $\pm$$1\,\sigma$$^{(2)}$ &  RV$_{lens}$ & $\pm$$1\,\sigma$$^{(2)}$ &  $\Delta$RV & $\pm$$1\,\sigma$$^{(2)}$  \\
-2\,456\,000 & km\,s$^{-1}$ & km\,s$^{-1}$& km\,s$^{-1}$& km\,s$^{-1}$ & km\,s$^{-1}$ & km\,s$^{-1}$ \\
\hline                        
   573.52363 & -13.166 &  0.199 &29.149  & 0.164 &42.315 & 0.076\\   
   721.88561 & -13.606  & 0.047 &28.762 &  0.063 &42.369 & 0.052  \\
   722.85366 & -13.011  & 0.025 &29.426  & 0.058 &42.437 & 0.053   \\
   729.83683 & -13.489 &  0.195 &28.895 & 0.161  &42.384 & 0.064  \\
   746.84698  & -13.543 &  0.051 &29.107 &  0.103 &42.650 & 0.109 \\ 
   769.80834 & -13.196  & 0.247 & 29.155 &  0.197 &42.351 &0.082   \\
   771.82660 & -12.779  & 0.162 & 29.678 &  0.150 &42.458 &0.060  \\
   830.73591  & -13.439 &  0.024 &29.030 &  0.043 &42.469 & 0.045 \\ 
   864.62468 & -13.140 &  0.132 & 29.223  & 0.112 &42.363 &0.049  \\
   925.52640 & -13.074 &  0.070 & 29.449 &  0.064 &42.522 &0.059  \\
\hline                                   
\end{tabular}
\tablefoot{(1) The BJD are UTC. (2) The 1-$\sigma$ error take into account the estimated photon-noise and the error due to the drift of the instrument (in the text $\sigma_{\mathrm pn}$ and $\sigma_{\mathrm calib}$, respectively).}
\end{table}
%
%

%
\begin{sidewaystable*}
\caption{Log of the observations. All RV measurements are kept in the analysis because weather degradations did not induce significant RV changes (see text). BJD, seeing and airmass values are given at mid-exposure.}             
\label{table_log}      
\centering          
\begin{tabular}{c c c c c c c c l l }     
\hline\hline        
Date  & BJD  & BERV & RV$_{\mathrm O{\tiny 2}}$$^{(1)}$ & SNR$^{(2)}$ & Mid-exposure$^{(3)}$ & Texp &  Airmass & Seeing &Comments \\ 
    & -2\,400\,000 & km\,s$^{-1}$ & km\,s$^{-1}$  & & & sec & & & \\
\hline                    
   2013-10-07 & 56573.52363 &       -27.94246	         & 	1.1275    	& 18.6		        & 0.44& 3480&  1.46& 1.09 & Seeing+10\% \\  
   2014-03-04 & 56721.88561 & 	29.70781	        & 	0.2125   	&    20.6		& 0.54&  3480& 1.16 &1.01 & Target drifted out of the slit \\
   2014-03-05 & 56722.85366 & 	29.83478	& 	0.9318   	&    22.0		& 0.51& 3480& 1.30  &0.75&   \\
   2014-03-12 & 56729.83683 & 	30.14055	& 	-0.6964  	&    21.6		& 0.44& 3480& 1.28& 0.69  &  \\
   2014-03-29 & 56746.84698  & 	29.02487	& 	-0.0933  	&    13.5		& 0.51& 3080&1.07& 1.76  &  Seeing deteriorated to the point    \\
  	              &                          &                               &                               &                               &           &        &       &    &  that the OB was ended. Wind was high too          \\
   2014-04-21 & 56769.80834  & 	23.79164	& 	-0.1324   	&   19.5		& 0.50& 3480 & 1.03&  0.60 &  \\  
   2014-04-23 & 56771.82660  & 	23.10472	& 	0.4227   	&   21.0		& 0.49 & 3600 & 1.01& 1.01  &  \\
   2014-06-21 & 56830.73591  & 	-3.24235	&      0.9928        	&    24.0		& 0.50& 3600 & 1.07&  1.00 &  \\
   2014-07-25 & 56864.62468 & 	-18.49718	& 	-0.0896  	&  24.0		& 0.52 & 3600 &  1.04& 0.71  &  \\
   2014-09-24 & 56925.52640 & 	-29.52330	& 	-0.4456  	&  21.6		& 0.48& 3600 & 1.24& 0.74  & Guide probe around 0.8"-1.0" \\
\hline                  
\end{tabular}
\tablefoot{(1) RV$_{\mathrm O{\tiny 2}}$ is the RV derived from the cross-correlation of the spectra with a telluric O$_{\mathrm 2}$ mask. (2) The SNR is per pixel at $\sim$550~nm. (3) The mid-exposure is calculated from the photometer count of the detectors and weighting by the spectral information between the blue and the red arms. }
\end{sidewaystable*}
%

\end{document}